\documentclass{edm_article}

\begin{document}

%\title{A Sample {\ttlit EDM} Proceedings Paper in LaTeX
%Format\titlenote{(Does NOT produce the permission block, copyright information nor page numbering). For use with ACM\_PROC\_ARTICLE-SP.CLS. Supported by ACM.}}
\title{Automatic Assessment of the Design Quality of Python Programs with Personalized Feedback}

%\subtitle{[Extended Abstract]
%\titlenote{A full version of this paper is available as
%\textit{Author's Guide to Preparing ACM SIG Proceedings Using
%\LaTeX$2_\epsilon$\ and BibTeX} at
% Submissions for EDM are double-blind: please do not include any
% author names or affiliations in the submission. 
% Anonymous authors:
\numberofauthors{2}
\author{
J. Walker Orr\\
       \affaddr{George Fox University}\\
       \email{jorr@georgefox.edu}
       \and
Nathaniel Russell\\
       \affaddr{George Fox University}\\
       \email{nrussell18@georgefox.edu}
}

\maketitle

%\onecolumn

%NOTE the reference section does not count towards the 6 page limit

\begin{abstract}
The assessment of program functionality can generally be accomplished with straight-forward unit tests.
However, assessing the design quality of a program is a much more difficult and nuanced problem.
Design quality is an important consideration since it affects the readability and maintainability of programs.
Assessing design quality and giving personalized feedback is very time consuming task for instructors and teaching assistants.
This limits the scale of giving personalized feedback to small class settings.
Further, design quality is nuanced and is difficult to concisely express as a set of rules.
For these reasons, we propose a neural network model to both automatically assess the design of a program and provide personalized feedback to guide students on how to make corrections.
The model's effectiveness is evaluated on a corpus of student programs written in Python.
The model has an accuracy rate from 83.67\% to 94.27\%, depending on the dataset, when predicting design scores as compared to historical instructor assessment.
Finally, we present a study where students tried to improve the design of their programs based on the personalized feedback produced by the model.
Students who participated in the study improved their program design scores by 19.58\%.
\end{abstract}

%% A category with the (minimum) three required fields
%%A category including the fourth, optional field follows...
%\category{D.2.8}{Software Engineering}{Metrics}[complexity measures, performance measures]

 % NOT required for Proceedings

\section{Introduction}
Recently there has a been a lot of work in the development of tools for education in programming and computer science.
Specifically there are many systems for intelligent tutoring which are designed to help students learn how to solve a programming challenge.
The tutoring involved is primarily focused in suggesting functional improvements, that is, how to finish the program so that it works correctly.

%intelligent tutoring, developing models to make improvements
Intelligent tutors such as \cite{rlhint} uses reinforcement learning to predict a useful hint in the form of an edit to a student's program that will get them one step closer to the goal of a functioning program.
It uses histories of edits made by students, starting with a blank slate and ultimately terminating with a functional program to train the model.
The system is based on \textit{Continuous Hint Factory} \cite{conthint} which uses a regression function to predict a vector that represents the best hint then translates that vector into a human-readable edit. 
Similarly \cite{progembed} used a neural network to embed programs and predict the program output.
Using that model of the program output, an algorithm was developed to provide feedback to the student on how to correct their program.
Also, \cite{studentknowledge} use a recurrent neural network to predict student success at a task given a history of student submissions of their program for evaluations.
%TODO add deepgrade reference and writeup here

All these systems model student programs from \textit{Hour of Code} \cite{hoc}.
\textit{Hour of Code} is a massively open online course platform that teaches people how to code with a visual programming language.
The language is simple and does not contain control constructs such as \textit{loops}.

%these tools are effectively designed to perform auto assessment of the 
%functionality of programs
Moreover, the combination of language and problem setting are simple enough that there is a single or very few functional solutions for each problem \cite{rlhint}.
This level of simplicity precludes the consideration of program design.
However for general purpose programming languages such as Python, there are many ways of creating functionally equivalent programs.
It is important for the sake of maintainability, modularity, clarity, and re-usability that students learn how to design programs well.

%talk about design
When it comes to the quality of design, there are varying standards.
Further, some standards are more objective or easier to precisely identify that others.
For example, the use of global variables are both widely recognized as poor design and are easy to identify.
%talk about pylint
For some programming languages, ``linters'' exist to apply rules to check for common design flaws.
For Python, Pylint \cite{pylint} is a code analysis tool to detect common violations of good software design.
It detects design problems such as the use of global variables, functions that are too long or take too many arguments, and functions that use too many variables.
Pylint is design to enforce the official standards of the Python programming community codified in PEP 8 \cite{pep8}.

%difficulty of using simple rules to capture good design qualities
%it is visible and able to be assessed to an instructor, but it is difficult to
%directly quantify
There are aspects of good design that are difficult to identify.
For example, simple logic is a good design idea, but is quite nebulous.
The complexity of a program's logic is contextual, it entirely depends on the problem the program is solving.
Also, modularity is universally judged as a quality of good design, however it is not always clear to what extent a program should be made modular.
How many functions or classes are too many?
Again, it depends on the context of the problem for which the program is designed.

%talk about software quality evaluation systems
In a professional setting, code reviews are often practiced to promote quality design that goes beyond the straight-forward rules of ``linters.''
Code reviews are a manual process which require a lot of human effort.
A recently developed system call DeepCodeReviewer \cite{deepcodereview} automates the code review process with a deep learning model.
By using proprietary data on historical code reviews taken from a Microsoft software version control system, DeepCodeReviewer was trained to successfully annotate segments of C\# code with useful comments on the code's quality.

%TODO talk about the problems: time consuming, does not scale up to MOOCs or even large classes, consistency, needs to be tailored to instructor needs
However, to our knowledge, there is no system to perform in-depth code analysis for the purposes of evaluating and assessing design for general purpose languages in an educational context.
The process of assessing the design of a program is time consuming for instructors and teaching assistants and it is an important component of complete intelligent tutoring system.
Such a system needs to be adjusted or calibrated for the context of particular problems or assignments since there are important aspects of software design are context dependent.
Moreover the system needs to match the particular standards of an instructor.
Hence we propose a system that models design quality with a neural network trained on previously assessed programs.
%TODO point out that training means contextualization, limited needed data means flexibility
%TODO talk a bit more about automatic assessment

%TODO an automatic system could be used to create an intelligent tutor for design quality, incorporated into a MOOC

\subsection{Our Approach and Contribution}

%TODO point out that limited need for training data means not 
%describe the setting and approach
We propose a design quality assessment system based on a feed-forward neural network that utilizes an abstract syntax tree (AST) to represent programs.
The neural network is a regression model that is trained on assessed student programs to predict a score between zero and one.
Each feature the model uses is designed to be meaningful to human interpretation and is based on statistics collected from the program's AST.
We intentionally do not use deep learning as it would make the representation of the program difficult to understand.
Personalized feedback is generated based on each feature of an individual program.
By swapping a feature's value for an individual program with the average feature value of good programs, it is possible to determine which changes need to be made to the program to improve is design.
The primary contributions of this work are the following:

\begin{itemize}

    %first to explicitly predict design quality in an educational setting
    \item The first to explicitly predict the design quality of programs in an educational setting to the best of our knowledge.

    %highly effective approach that works with limited data
    \item High efficacy with an accuracy from 83.67\% to 94.27\% with only small amounts of training data required.
    
    \item The first intelligent tutoring system for design quality for Python.

    %no need for explicit training for feedback    
    \item Personalized feedback without the explicit training or annotation.

%TODO tutoring and assessment on Python rather than block language
%TODO no need for trajectories of program edits

\end{itemize}

%TODO finish - is just handled by the intro?
%\section{Related Work}
%Lorem ipsum dolor sit amet, consectetur adipiscing elit, sed do eiusmod tempor incididunt ut labore et dolore magna aliqua. Ut enim ad minim veniam, quis nostrud exercitation ullamco laboris nisi ut aliquip ex ea commodo consequat. Duis aute irure dolor in reprehenderit in voluptate velit esse cillum dolore eu fugiat nulla pariatur. Excepteur sint occaecat cupidatat non proident, sunt in culpa qui officia deserunt mollit anim id est laborum. 

%TODO finish
%\section{Problem}
%Lorem ipsum dolor sit amet, consectetur adipiscing elit, sed do eiusmod tempor incididunt ut labore et dolore magna aliqua. Ut enim ad minim veniam, quis nostrud exercitation ullamco laboris nisi ut aliquip ex ea commodo consequat. Duis aute irure dolor in reprehenderit in voluptate velit esse cillum dolore eu fugiat nulla pariatur. Excepteur sint occaecat cupidatat non proident, sunt in culpa qui officia deserunt mollit anim id est laborum. 

%TODO describe setup, python program, design score

%TODO manual feedback in the form of comments

\section{Method}
\begin{figure}
    \centering
    \includegraphics[scale=.6]{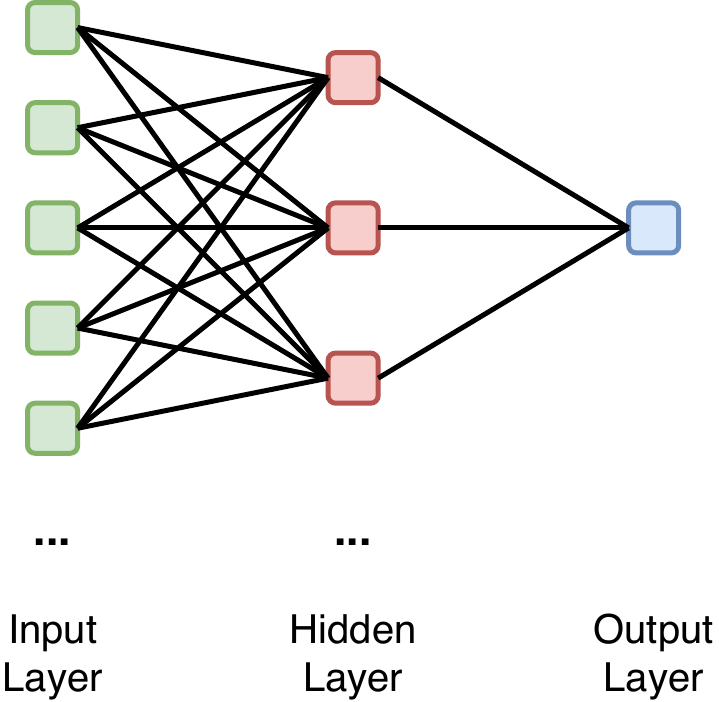}
    \caption{The model of program design quality, a feed-forward neural network. The ``Input Layer'' is the feature vector created from the AST. The ``Hidden Layer'' corresponds to calculation of $x^\prime$ specified in Equation \ref{eq:hidden}. Finally, the ``Output Layer'' produces a single value, the design score as found in Equation \ref{eq:out}.}
    \label{fig:model}
\end{figure}

The task is to predict a design quality score for a student program written in Python.
The score $y$ is a real number between zero and one.
The program is represented by a feature vector $\vec{x}$ produced by the output of a series of feature functions computed from the program's AST.

%overview of approach feature extraction, NN to predict score
For an AST $T$, a series of feature functions $f_i((T)$ output is concatenated in to a feature vector $\vec{x}$ that represents key aspects of the program's design.
The model $g(\vec{x}; \Theta)$ is a feed-forward neural network with a single hidden layer.
It is a regression model that predicts the score $y$ based on the feature vector $\vec{x}$ and parameters $\Theta$.

%TODO using average feature value to recommend which feature to change, which means 

%TODO finish
\subsection{Features}

\begin{figure*}
    \centering
    \includegraphics[scale=.35]{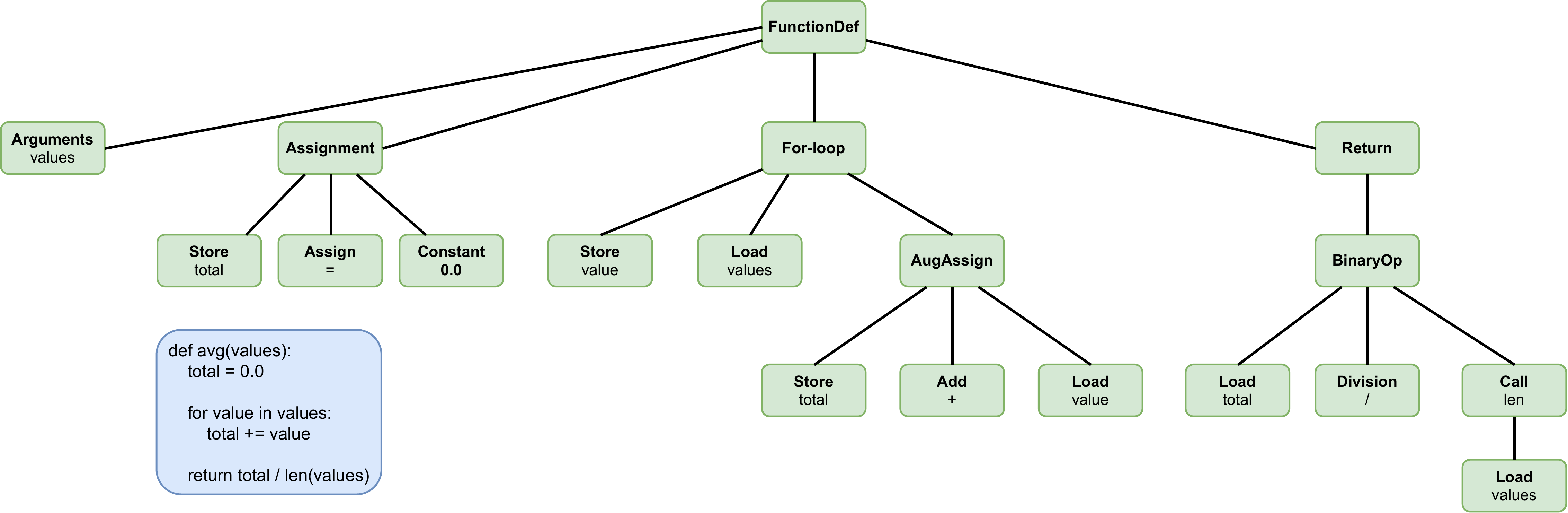}
    \caption{An abstract syntax tree for a segment of Python code that computes the average of the values in a list. The AST has been condensed for the sake of brevity and the restrictions of space. The related code segment is shown in blue.}
    \label{fig:ast}
\end{figure*}

Despite recent advances in deep learning, we chose to represent the student program with feature functions computed on its AST.
Deep learning is highly effective at learning useful feature representations of everything from images to time series to natural language texts.
However, deep learning also requires large amounts of data and in this setting the quantity of manual annotated student programs is limited.

Additionally, AST are a natural and effective means of representing and understanding programs and can be created with free, available tools. 
An AST is an exact representation of the source code of program based on the programming language's grammatical structure.
Producing an AST representation of a programming language is an essential first step in compilers and interpreters.
The AST of a program contains all the content of its source but also is augmented with the syntactic relationships between every element.
A parser and tokenizer to produce an AST for the Python programming language is provided by its own standard library.
This makes the AST the natural representation to use, since it is free, convenient, exact, interpretable, and does not required any additional data.
In contrast, deep learning would require a large amount of data to effectively reproduce the same representation.

Prior to representation as an AST, a program must first be broken into a series of tokens via the process of lexicalization.
Lexicalization is the process of reading a program, character-by-character and dividing into work-like tokens.
These tokens are also assigned a type such as a function call or variable reference.
The grammatical rules of the language are applied to the lexicalized program to create the AST.
In an AST, the leaf nodes of the tree are the program's tokens while the interior nodes correspond to syntactic elements and constructions.
For example, an interior node could represent the body of a function or the assignment of a value to a variable.
An example of an AST is found in Figure \ref{fig:ast}.

Given that an AST is a complete representation of a program, it is a natural basis for assessing the quality of a program's design.
Deep learning may be able to automatically learning the same key syntactic relationships with enough data, however this information is simply available via AST.
Further, features computed from the AST will be human interpretable unlike a representation produced by deep learning.

The features we created are all based on statistics collected from a program's AST.
Some consist of simply counting the number of nodes of a given type, for example, the number of user defined functions.
Other feature functions are based of subsections of the AST, such as the number of nodes per line or per function.
Finally, some features are ratios or percentages such as the average percent of lines in a number in a function that are empty.
All of the features are relatively simple and fast to compute, yet generally capture the design and quality of a program.
Each of the feature functions $f_i(T)$ we defined can be found in Appendix \ref{appendix:ff}.

\subsection{Model}

The model is a feed-forward neural network \cite{mlp} with a single hidden layer and single neuron in the output layer.
The model's structure is illustrated in Figure \ref{fig:model}.
The values of the input layer are the feature vector $\vec{x}$.
Each neuron in the hidden layer $x^\prime_j$ defined with the following equation:
\begin{align}
    x^\prime_j = \text{ReLU}\bigg(\sum\limits_{i = 1}^{d} w_{i,j} x_i\bigg) \label{eq:hidden}
\end{align}

where $d$ is the dimension of $\vec{x}$ and $w_{i,j} \in \Theta$ are the parameters, ``weights'' of the neuron.
We use the ReLU \cite{relu} as the activation function for the hidden layer neurons.
The final prediction of the design score is made by the output layer's single neuron:
\begin{align}
    y = \sigma\bigg(\sum\limits_{j = 1}^{d^\prime} w_{j} x^{\prime}_j \bigg) \label{eq:out}
\end{align}

where $d^\prime$ is the number of hidden layer neurons and $w_j \in \Theta$ are weights of this neuron.
The function sigmoid is used because its domain spans from $(-\infty, \infty)$ but its range is $[0, 1]$ which ultimately guarantees the model always outputs a valid score.
The model is trained with mean squared error as the loss function:
\begin{align}
    \text{MSE} = \frac{1}{n} \sum \limits_{k = 1}^{n} ( y^*_k - y_k )^2 
\end{align}

where $y^*_k$ is the ground-truth design score for the $k^{\text{th}}$ instance i.e. program and $n$ is the number of instances in the training data.
The model is trained with the ADAM algorithm \cite{adam} and each parameter in the model was regularized according to their L2 norm \cite{l2norm}.
For all our experiments, a hidden layer of size 32 was used.
The model was trained for 250 epochs and the model from the best round according to a development set was selected for our experiments.
%model settings, 32 hidden, epochs 250, best after 10

\subsection{Ensemble}
Due to the fact that fitting neural networks to data is a local optimization problem, the effect of initial values of the parameters $\Theta$ of the model remain after training.
The process of training a neural network will produce a different model given the same data.
This variation in the results of a trained model is particularly pronounced when the training data set is relatively small.
To address this variation and mitigate its impact an ensemble of models can trained, each with different initial parameter values.
Each model is independently trained and a single prediction is made by a simple of the average of individual predictions i.e.
\begin{align}
    y = \frac{1}{m} \sum \limits_{l = 1}^{m} y_l 
\end{align}

where $m$ is the number of models and $y_l$ is the prediction of the $l^{th}$ model.
For our experiments, an ensemble of 10 models was used.
%TODO write more?

\subsection{Personalized Feedback}
\label{sec:feedback}
The goal of intelligent tutors is to provide personalized feedback and suggestions on how to improve a program.
The most straight-forward means of providing feedback would be to simply predict which possible improvements apply to a given program.
However, training a model to directly predict relevant feedback would require a dataset of program with corresponding feedback and such a dataset can be hard to find or is expense to construct.

In order to avoid the need for a dataset with explicit feedback annotation, we use our model, trained on predicting design score, to evaluate how changes in program features would lead to a higher assessed score.
Using the training data, we compute an average feature vector $\underline{\vec{x}}$ of all the ``good'' programs i.e. those with a design score greater than 0.75.
To generate feedback for a program, its feature vector $\vec{x}$ is compared to the average $\underline{\vec{x}}$.
For each feature, a new vector $\vec{x}^\prime$ is created by replacing the feature value $x_i$ with value with the average's value $\underline{x}_i$.
This process is setting up a hypothesis, what if the program was closer to the average ``good'' program with regards to a particular feature?
To answer this, the trained model $g(\vec{x}; \vec{\theta})$ is used to predict a design score for the new vector i.e. $y^\prime_i = g(\vec{x}^\prime; \vec{\theta})$.
By comparing the original score of the program $y$ with the new score $y^\prime_i$, the hypothesis can be tested.
If the new score $y^\prime_i$ is greater than the original predicted score $y$, then the alteration of $x_i$ to be closer to $\underline{x}_i$ is an improvement.
Feedback based on this alteration is recommended to the student as personalized feedback.
Since each feature in $\vec{x}$ is understandable to a human, feedback is given in the form of the suggestion to increase or decrease particular features.
The suggestion for alteration is based on the comparison of $x_i$ versus $\underline{x}_i$, if $x_i > \underline{x}_i$, the feedback of decrease $x_i$ is given. 
In the other case, where $x_i < \underline{x}_i$ the feedback is to increase $x_i$.
Based on the feature and the feedback of increase or decrease, a user-friendly sentence is selected from a table of predefined responses.
For example, if $x_i$ is the number of user defined functions and $x_i = 3$, $\underline{x} = 5$, and $y^\prime_i > y$ then the feedback of ``increase the number of user defined functions'' is created.

%TODO redo combined
\section{Experiments}
\begin{table*}[]
    \centering
    \begin{tabular}{| l | l  l | l  l | l  l | l  l | l  l |}
        \hline
        \multicolumn{1}{|c}{\textbf{Method}}
        & \multicolumn{2}{ | c}{\textbf{Travel}}
        & \multicolumn{2}{ | c}{\textbf{Budget}}
        & \multicolumn{2}{ | c}{\textbf{RPS}} 
        & \multicolumn{2}{| c}{\textbf{Craps}} 
        & \multicolumn{2}{| c |}{\textbf{Combined}}\\
         & MSE & Accuracy & MSE & Accuracy & MSE & Accuracy & MSE & Accuracy & MSE & Accuracy\\
        \hline
        Linear Regression & 0.009 & 93.09\% & 0.032 & 87.43\% & 0.038 & 83.2\% & \textbf{0.043} & \textbf{84.06\%} & \textbf{0.027} & \textbf{87.03\%} \\
        Decision Tree & 0.018 & 90.10\% & 0.031 & 87.26\% & 0.078 & 77.33\% & 0.072 & 79.41\% & 0.076 & 81.42\%\\
        Sig. Linear Regression & 0.022 & 89.60\% & 0.046 & 85.13\% & 0.086 & 77.91\% & 0.063 & 81.43\% & 0.070 & 80.64\% \\
        Neural Network & 0.007 & 93.48\% & 0.024 & 88.48\% & 0.041 & 83.9\% & 0.08 & 79.57\% & 0.033 & 85.61\%\\
        Ensemble & \textbf{0.005} & \textbf{94.27\%} & \textbf{0.022} & \textbf{90.14\%}  & \textbf{0.022} & \textbf{87.66\%} & 0.053 & 83.67\% & 0.031 & 86.99\%\\
        \hline
    \end{tabular}
    \caption{Design Score Prediction Results}
    \label{tab:results}
\end{table*}

%two types of evaluation: prediction accuracy and case study
%two small datasets of programs of historical records
The system was evaluated in two different experimental settings.
The first evaluation is direct test of the model's accuracy on known design scores.
For this, several different datasets and settings were compared against several baselines.
The second evaluation is a small study of how student's responded to the system's feedback.
%case study data related to RPS
Students were given feedback on the quality of their programs based on the model.
They were given the chance to correct their programs after receiving feedback and have it manually reassessed.

\subsection{Dataset}

%described dataset
The dataset was collected over three years from an introduction to computer science course which teaches Python 3.
It consists of four separate programming assignments which involve a wide range of programming skills.
The simplest is ``Travel,'' an assignment that involves the distance a vehicle travelled after going a constant speed for a specified duration.
There are 118 student programs for ``Travel.''
The next assignment, ``Budget'' is a budgeting program that lets a user specify a budget and expenses and determines if they are over or under their budget.
168 student programs were collected for ``Budget''.
The third assignment in the dataset consists of creating a program to play ``Rock-Paper-Scissors'' against the computer.
For this assignment, there are 111 student programs.
The last assignment is programming the classic casino game ``Craps'' which involves rolling multiple dice and placing different types of bets and wagers.
This assignment has 120 collected student programs.

All the assignments require the student to write the program from scratch in Python 3.
The programs are to have a command-line, text-based interface and user validation.
%play multiple rounds versus the computer with an overall score across all the games.
Students are required to use if-statements, loops, user defined functions.
The ``Craps'' program also requires the student to do exception handling, and file I/O.
A requirement of the program was to maintain a record of their winnings across sessions of playing the game, hence the results were required to be stored to a file.
Also, the standards of design quality go up as the course progresses and since ``Craps'' is the last assignment, it has the highest standards.
Each student program has an associated design score that was normalized to value between zero and one.

\subsection{Baseline Methods}

The model is compared against a variety of baseline regression methods.
The simplest is linear regression, which simply learns a weight per each feature.
Next is a regression decision tree which is trained with the CART algorithm \cite{cart}.
It has the advantage over linear regression in that it can learn non-linear relationships.
Non-linearity means a model can learn ``sweet-spots'' rather than simply having a ``more is better'' understanding of some features.
For example, having some modularity in the form of user defined functions is good, however, too many is cumbersome.
The ``correct'' number of user defined functions likely should fall into a relatively small range.
Model selection on the maximum depth of the tree with a development set was used to determine that 10 was the best setting.

However, both of these models have the issue that they are not constrained to produce a score between zero and one, their prediction can be any real number.
Hence another baseline method was used, created to be an intermediary step between linear regression and the neural network model.
It is a linear model with a sigmoid transformation which guarantees the output be between zero and one.
This model is effectively the final layer of the neural network model, i.e. the neural network without the hidden layer.
The model is specified by the equation:
\begin{align}
    y = \sigma\bigg( \sum\limits_{i = 1}^{d} w_{i,j} x_i \bigg) 
\end{align}

This model is also trained with ADAM \cite{adam}.
All the baseline models and the neural network are trained with MSE as the loss function.

\subsection{Results}

%define "accuracy"
The model was compared versus each baseline in five different settings: the ``Travel'', ``Budget'', ``Rock-Paper-Scissors'', and ``Craps'' programs, and a combined dataset which includes all the programs.
The results of the experiments can be found in Table \ref{tab:results}.
Each model is evaluated according to two different metrics: MSE and average accuracy.
Average accuracy is defined as $\frac{1}{n} \sum \limits_{k=1}^n \big( 1 - |y_k - y^*_k| \big)$.

%compare each method
%compare ensemble
Overall, the decision tree and the sigmoid-transformed were clearly the two worst models.
This was surprising since decision trees are generally thought to be strictly more powerful than linear models.
However, decision trees look for highly discriminative features to partition the data into more consistent groups.
The under-performance of the decision tree possibly indicates that none of the features were especially indicative of a good or bad design on their own.
Instead, the quality of a program is better described by a collection of subtle features, which gives credence to the belief that design quality is nuanced.

The reason sigmoid-transformed linear model under-performed linear regression was likely due to it being trained with ADAM.
ADAM does not guarantee convergence to a global optimum like the analytical solution to linear regression.
Apparently the restriction on predictions to be within the specified range of zero to one was not important.

Linear regression did surprisingly well, beating both the neural network and network ensemble in the ``Craps'' and combined datasets, though barely.
In those cases, the difference between the ensemble and linear regression was less than a percent.
This is likely due to the stability of linear regression's predictions.
Though linear regression does not have the power and flexibility of neural networks this can also be a benefit by limiting how wrong their predictions are.
Neural networks and even ensembles can make overconfident predictions on outliers or other unusual cases.
The ``Craps'' dataset contained the most complex programs and it is likely a handful of predictions significantly brought down the average.

The network ensemble outperformed the single neural network in every case, which is to be expected.
The margin of improvement of the ensemble versus the single neural network in accuracy on the four individual program datasets ranged from ~1\% to ~4\%.
The network ensemble did the best overall by being the best in most cases or coming in a close second in all the other cases.
The importance of using an ensemble is evident on the ``Craps'' dataset where the individual neural network under-performed significantly.
On the other datasets, the neural network outperformed linear regression by a small margin, but on ``Craps'' the neural network model under-performed the linear regression model by 5\%.
Again, this is most likely due the instability and variability of neural network predictions i.e. small differences in features can lead to a large difference in the prediction.
In the ``Craps'' dataset, the improvement of the ensemble over the single neural network model illustrates the relative stability of the ensemble's predictions.
In every case, the ensemble is superior to the single neural network and had the best overall performance by producing the most accurate results on three of the datasets and effectively tying for the best on the other two.

%need for individual models
One noticeable pattern was that all the models performed better on the ``Travel'' and ``Budget'' datasets than on the ``RPS'', ``Craps'', and combined datasets.
Universally, the most difficult dataset was ``Craps'' which likely lowers the accuracy on the combined dataset.
Due to the shifting standards and expectations of student assignments, a model per assignment appears to worthwhile.
This is a bit counter-intuitive since there are many common standards and expectations across assignments.

%practicality of the system
Overall, the network ensemble produced reliable, accurate results when trained per dataset.
The accuracy of the ensemble is arguably close to being useful in practical application.
Further, comparing the scores of an instructor versus another instructor or even against themselves, the rate of agreement must be less than 100\% and with an accuracy of the network ensemble ranging from 83.67\% to 94.27\%, the model's accuracy is possibly close to a realistic ceiling.

%TODO add more...

%TODO finish - include table or figure?
\subsection{Feedback Study}

In order the evaluate the effectiveness of the personalized feedback, we conducted a small study on the effect of the feedback on the design score of student programs.
For the ``Rock-Paper-Scissors'' program the network ensemble was used to generate personalized feedback for the student programs instead of the usual instructor feedback.
The network ensemble was the same as used in the design score experiments, it was trained with prior years worth of student programs.
Having received the personalized feedback, students opted into correcting their program for extra credit on their assignment.
The feedback was in form of a series of comments, where each comment was ``increase'' or ``decrease'' the name of a feature as described in Section \ref{sec:feedback}.

%collect data
The class is an introduction to computer science course with multiple sections and two different instructors.
Students from both instructors participated in the study.
Out of 73 students enrolled across the sections of the course, 15 students chose to opt-in.

%average improvement
The revised programs were assessed again manually for design quality and the scores were compared against the originals.
The design score of the programs started at an average of 68.33\% and after the feedback and correction the average rose to 87.92\%, a 19.58\% absolute improvement.
Using a paired t-test, the improvement was judged to be significant with a p-value of 0.001.

%admit to caveats: small size, single assessor, opt-in
The results of the study suggest the feedback was generally useful in guiding students to improve the design quality of their programs.
The improvement was noticeable to the instructors anecdotally as well.
For example, the usage of global variables and ``magic numbers'' decreased significantly.
Though the study does have some caveats including its small sample size and opt-in participation.
It could be that those students willing to opt-in are those most willing or able improve with a second chance.

%TODO finish
\section{Conclusions \& Future Work}

Overall, we proposed a neural network model ensemble for predict the design quality score of a student program and experimentally demonstrated its effectiveness.
Further, our system provided personalized feedback based on the difference between a program's feature values and the average features' value of ``good'' programs.
A small study provides evidence that the feedback was of practical use to students.
Students were able to improve their programs significantly based on the feedback they received.

% transfer learning across assignments
% more data, more programs evaluated
There is also evidence that training models per assignment is most effective.
However, the model needs to be evaluated on more programming assignments.
Further, there is a possibility of utilizing transfer learning \cite{transferlearning} to help the model learn what is in common across the assignments.

% explicit training for feedback - magic numbers etc
% more nuanced feedback - per line, could use deep learning
% multi-instance learning
% active learning?
The feedback given was shown to be effective, but more nuanced feedback could be useful.
Specifically, feedback targeted to individual lines or segments of code would possibly help students improve their program's more effectively.
However, this may require additional supervision i.e. annotation for explicit training.
Active learning \cite{activelearning} or multi-instance learning \cite{multiinstancelearning} may be alternatives to gathering additional annotation.

% intelligent tutor
%Finally, this model would be 

%\end{document}  % This is where a 'short' article might terminate

%ACKNOWLEDGMENTS are optional
%\section{Acknowledgments}
%Lorem ipsum dolor sit amet, consectetur adipiscing elit, sed do eiusmod tempor incididunt ut labore et dolore magna aliqua. Ut enim ad minim veniam, quis nostrud exercitation ullamco laboris nisi ut aliquip ex ea commodo consequat. Duis aute irure dolor in reprehenderit in voluptate velit esse cillum dolore eu fugiat nulla pariatur. Excepteur sint occaecat cupidatat non proident, sunt in culpa qui officia deserunt mollit anim id est laborum. 

%TODO glory to God alone

\bibliographystyle{abbrv}
\bibliography{main}  % sigproc.bib is the name of the Bibliography in this 

\appendix

\section{Feature Functions}
\label{appendix:ff}

\begin{itemize}
    \item The number of functions
    \item The number of assignments
    \item AST nodes per function
    \item Lines of code per function
    \item Total lines of code
    \item Number of literals
    \item The proportion of white-space characters to the total number of characters
    \item Number of empty lines
    \item Deepest level of indentation
    \item Number of ``if'' statements
    \item Number of comments
    \item Number of AST nodes per lines of code
    \item Number of try-except statements
    \item AST nodes per try-except statement
    \item AST nodes per ``if'' statement
    \item Number of lists
    \item Number of tuples
    \item Average line number of literals
    \item Average line number of function definition
    \item Average line number of ``if'' statement
    \item Ratio of AST nodes inside functions versus total number of AST nodes
    \item Number of function calls
    \item Number of ``pass'' statements
    \item Number of ``break'' statements
    \item Number of ``continue'' statements
    \item Number of global variables
    \item Number of zero and one integer literals
    \item Average line number of ``import'' statement
    \item Number of numeric literals
    \item Number of comparisons
    \item Number of ``return'' statements
    \item Maximum number of ``return'' statements per function
    \item Maximum number of literals per ``if'' statement
\end{itemize}

\balancecolumns
% That's all folks!
\end{document}